# Atomic scale control and visualization of topological quantum phase transition in π-conjugated polymers driven by their length

H. Gonzalez Herrero[1†], J. Mendieta-Moreno[2], Sh. Edalatmanesh[1,2], J. Santos[3,4], N. Martín[3,4], D. Écija[4], B. de la Torre[1,2], P. Jelinek[1,2*]

**Affiliations:**

[1] Regional Centre of Advanced Technologies and Materials, Czech Advanced Technology and Research Institute, Palacký University, 78371 Olomouc, Czech Republic.

[2] Institute of Physics, Czech Academy of Sciences, 162 00 Prague, Czech Republic.

[3] Department of Organic Chemistry, Faculty of Chemistry, University Complutense of Madrid, 28040 Madrid, Spain.

[4] IMDEA-Nanociencia, 28049 Madrid, Spain

*Corresponding author. Email: jelinekp@fzu.cz (P.J.).

†Present address: Department of Applied Physics, Aalto University, 02150 Espoo, Finland.

**Abstract:** Quantum phase transitions, which are driven by quantum fluctuations, mark a frontier between distinct quantum phases of matter. However, our understanding and control of such phenomena is still limited. Here we report an atomic scale control of quantum phase transition between two different topological quantum classes of a well-defined π-conjugated polymer controlled by their length. We reveal that a pseudo Jahn-Teller effect is the driving mechanism of the phase transition, being activated above a certain polymer chain length. In addition, our theoretical calculations indicate the presence of long-time coherent fluctuations at finite temperature between the two quantum phases of the polymer near the phase transition. This work may pave new ways to achieve atomic scale control of quantum phase transitions, in particular in organic matter.

**One-Sentence Summary:** Atomic scale control and visualization of topological quantum phase transition in π -conjugated polymers activated by their length.

**Main Text:** In general, a phase transition represents a transformation process between two states of matter with distinct internal order involving long-range fluctuations across the entire system. Classical phase transitions are driven by thermal fluctuations, being well understood within the frame of Landau-Ginzburg-Wilson theory (1).

However, at zero temperature, where no thermal fluctuations occur, a quantum phase transition (QPT) (2) may happen, being driven by quantum fluctuations imposed by the principle of uncertainty, whenever a non-thermal parameter like pressure, magnetic field or chemical composition is modified (3, 4). Such quantum phase transitions are the subject of an intensive research nowadays in order to shed light into the fundamental mechanisms behind them, and due to their relevance in complex phenomena in solid state physics (5-7).

Typically, continuous QPTs take place in the proximity of the so-called avoided level crossing (see Fig. 1), whenever the energy difference between the ground state and the lowest excited state diminishes and becomes comparable to the quantum energy fluctuation. In principle, under such circumstances, only the quantum fluctuations may cause the phase transition into a new phase of matter at zero Kelvin (in the absence of any thermal fluctuation) (8, 9). Importantly, near the critical point at finite but very low temperature both thermal and quantum fluctuations are equally important, giving rise to the so-called quantum criticality phenomenon. The quantum critical state cannot be described within the conventional quasiparticle picture of the Fermi liquid theory (10, 11). This exotic quantum state plays significant role in our understanding of e.g. high temperature superconductivity (12-15), ferroelectricity (16, 17) or strange metals featuring non-Fermi liquid behavior (12). From this perspective, control and understanding of QPT represent one of the biggest challenges of contemporary science due to their complexity and the inherent quantum character (18). As a result, it is commonly perceived that a detailed knowledge may allow to design new quantum materials with unforeseen properties.

In condensed matter physics, a lot of the attention has been paid to QPT between phases featuring strongly correlated states of matter with complex spin and charge pattern frequently manifested by spin and charge density waves (CDW) (19, 20). However, it is precisely the presence of strongly correlated phases, which makes it significantly more difficult to gain a deeper understanding of the phase transition mechanism itself. Therefore, there is a quest to search for novel materials, where QPT takes place between two phases of matter with only moderated correlation, since such systems could be well described within mean field approximations. In addition, there is a growing interest in controlling the proximity to the quantum critical point by harnessing designer parameters with great precision, in order to enable the direct observation of the quantum phase transition.

In one-dimensional π-conjugated polymers, the Peierls instability (21) results in an alternation of bond lengths between carbon atoms accompanied by variation of corresponding bond charges, which gives rise to the π-conjugation (22, 23). The π-conjugation can also be viewed as CDW phase driven by a Fermi surface nesting and complex interplay between electronic and lattice degree of freedom (24, 25). Remarkably, even very simple polymers such as polyacetylene may feature distinct quantum topological phases as demonstrated by the celebrated SSH model (26), where one hallmark of the non-trivial topology of the electronic structure is the presence of in-gap energy edge states.

Recent progress in the field of on-surface synthesis (27, 28) supported by high-resolution imaging by scanning probe microscopy (29, 30) has brought a new chemical strategy to synthesize organic hydrocarbon materials with atomistic control (31) and inspect their inherent electronic, magnetic and topological properties (32-35). Following these on-surface synthetic protocols, in a recent study, we demonstrated that distinct π-conjugation resonance forms and corresponding topological phases can be tuned in acene-bridged π-conjugated polymers by a proper choice of acene units, i.e. the monomer constituents of the polymer (36). The change of quantum class from topologically trivial to non-trivial polymers was accompanied not only by significant reduction of the electronic band gap, but also by a rearrangement of π-bonds and the emergence of zero-energy edge states as a consequence of the non-trivial topology of the electronic band structure (37).

In this work, we reveal that the length of a 1D polymer can act as a non-thermal parameter to trigger a topological quantum phase transition. To exemplify experimentally this strategy, we employ pentacene bridged polymers (36) as archetypes. Such polymers belong to the topologically non-trivial quantum phase in the limit of infinite length. We carried out a systematic study of the evolution of the π-conjugation resonance form and level crossing of frontier orbitals (see schematic view in Fig. 1) accompanied by the emergence of the edge states upon increasing of polymer length by means of scanning probe microscopy (SPM), complemented by theoretical simulations. Our results reveal that the topological phase transition is driven by the length of the polymer. Such findings are rationalized by a multi-orbital pseudo Jahn-Teller effect (38) triggered by the length dependent vibronic coupling between frontier occupied and unoccupied orbitals. In addition, our density functional theory (DFT) molecular dynamics simulations indicate the presence of long-time coherent fluctuations at finite temperature between the two distinct π-conjugated/topological phases of the polymers situated near the topological phase transition.

Pentacene polymers (36, 39) were grown by deposition of precursor 4BrPn (6,13-bis(dibromomethylene)-6,13-dihydropentacene) on Au(111) surface under ultra-high vacuum conditions and subsequent annealing of the surface at 500K. This promotes homocoupling of the 4BrPn molecules leading to the formation of the pentacene polymers. In order to control the final length of our polymers, we intentionally varied the initial coverages of the precursor molecules. For coverages below 15%, finite polymers could be found (see fig. S1 for representative scanning tunneling microscopy (STM) images of 4%, 12% and 15% coverage, and the resulting polymers obtained after annealing). Following this protocol, we were able to produce a broad variety of pristine defect-free polymers with lengths ranging from 2 monomers up to ~50 monomers.

In a recent previous study (36), we investigated very long pentacene polymers (in range of 50-100 nm) grown on Au(111) showing the appearance of in-gap edge states as a consequence of the non-trivial topology of their electronic band structure. Herein, there is a question: if the length of a polymer could influence its topological quantum class. To explore such hypothesis, we employed the modified synthetic protocol, discovering the absence of in-gap edge states for chains shorter than 26 monomers. Such finding indicates that there is a critical length for the pentacene chains to exhibit the non-trivial topology. Figs. 2A-B display simultaneously acquired high-resolution atomic force microscopy (AFM) and STM images taken with CO-tip (29) of two different defect-free chains made of 24 and 26 monomers, respectively. We have deliberately chosen these two chains of a given length, for which a visible change in the electronic structure takes place. If we take a look to low bias STM image near the Fermi level (STM images at ~3 meV shown in Figs. 2A-B), we observe that the chain made of 26 monomers (and longer) features the in-gap edge states, while the chain of 24 monomers (and shorter) does not. The in-gap edge state is also presented in d$I$/d$V$ spectrum acquired at the end of the non-trivial chains as a conductance peak around the Fermi energy (see fig. S2).

Fig. 2C represents selected d$I$/d$V$ spectra of defect free polymers of different lengths with well-defined increase in the conductance corresponding to HOMO/LUMO orbitals, which determines the electronic band gap of the polymer. A systematic evolution of the HOMO/LUMO bandgap as function of their length is shown in Fig. 2D. As it can be seen, the bandgap of the 1D polymer varies continuously from ~1.2 eV for a dimer chain down to ~350meV (-150 meV for the HOMO energy and 200 meV for the LUMO) for the decamer. For longer polymers the value remains constant. However, the saturation of the bandgap does not seem to be directly related with the trivial/non-trivial transition of the polymers, since the QPT happens somewhere between 24 and 26 monomers. Fig. 2E also shows representative d$I$/d$V$ maps obtained at the HOMO/LUMO energies of the chains for different topological phases. Interestingly, the features observed for the HOMO orbital of trivial polymers are found in the LUMO orbital of the non-trivial chains and vice versa. Nonetheless, this behavior has been observed when comparing the CB/VB of anthracene vs pentacene polymers in previous works (36) and it is responsible for the change of Zak phase (40) and thus of the topological quantum class of the polymer. It is appropriated to note that Fig. 2D directly represents an avoided level crossing scheme associated with continuous quantum phase transition. This points out that the HOMO/LUMO crossing level plays a decisive role in the topological phase transition, as we will elaborate below.

The change of topological state of these polymers is also accompanied by a variation of the inherent π-conjugation, i.e. the resonance forms, going from ethynylene-aromatic (topologically trivial in acene polymers) to cumulene-quinoid (topologically non-trivial in pentacene polymers). Such a change in the character of the π-conjugation, considered as well as transition between two distinct CDW phases, can be directly observed in high-resolution AFM images obtained with CO functionalized tips. In Fig. 2F, we present two high-resolution AFM images of central part of trivial and non-trivial chains respectively, together with a profile along the line marked in red and blue. The presence of a protrusion in the bridging unit is associated with the triple bond of the ethynylene link of the topologically trivial chains (see red line profile in Fig. 2F). While the bright protrusions and corresponding peaks in the red line profile can be observed throughout the trivial polymer, the non-trivial chain shows a disappearance of the protrusions and a decay in intensity of the peaks as we get closer to the central part of the chain (see blue line profile in Fig. 2F). This variation demonstrates an ethynylene to cumulene-like transition along the topologically non-trivial polymer.

In order to improve the experimental capabilities of controlling the phase transition in π-conjugated polymers, we have developed a novel protocol employing atomic manipulation. Such strategy consists of two step process; hydrogenation of the pentacene polymers to afterwards selectively remove those extra hydrogen atoms from the desired part of the chain using the STM tip.

An exposure of pentacene polymers to atomic hydrogen in UHV chamber leads to highly selective adsorption of two hydrogens atoms on one of the benzene rings close to the central one (see Figs. 3A-B and details of hydrogenation process in Methods). Fig. 3B shows an experimental and theoretical AFM image of the hydrogen defects with excellent agreement between both images. Notably, such highly site-selective adsorption is dictated by the formation of a new π-resonance form with two Clar's sextets, with higher stabilization of the chemical structure. Therefore, the hydrogenation process increases substantially the electronic band gap (~1.8 eV see fig. S3) of the polymer establishing an ideal template for an artificial formation of large segments of pristine pentacene polymers of different lengths.

The hydrogen pair defects can be removed in a controlled way by placing the STM tip on top of one of them (a red dot in Fig. 3C) and performing a distance versus voltage curve, ramping the voltage from 1.5 V up to 2.5-3.0 V, while keeping a constant current (~10 pA). In the plot in Fig. 3C we can observe a sudden jump around 2.5 V pointing to a successful removal of the extra hydrogen atoms (see fig. S4 for snapshots of a hydrogen removal process). Following this protocol, we can selectively engineer regions of the pristine pentacene polymer with a desired topological phase, which depends on the length of such region. As an example, in Fig. 3D we show an AFM image of a long fully hydrogenated pentacene polymer before our manipulation experiment. The AFM image from Fig. 3E displays the chain after multiple manipulations (extra-hydrogen removals) where a trivial chain consisting of 16 monomers (left part of the chain) and a non-trivial chain made of 31 monomers (right part of the chain) have been created. The low bias STM image of the modified polymer shown in Fig. 3E reveals the existence of the edge state in the non-trivial part in opposition to the trivial chain. At the same time, d$I$/d$V$ maps acquired at the HOMO and LUMO energies (-150 meV and 200 meV respectively) prove once again the inversion of the orbitals between the trivial and non-trivial polymers (see two bottom panels of Fig. 3E). Importantly, the artificially formed pentacene polymers exhibit the same behavior as the finite pristine chains previously described.

To summarize the experimental observations, the quantum topological phase transition is accompanied by the avoided level crossing of frontier HOMO/LUMO orbitals and the transformation of π-conjugation resonance form, as shown schematically on Fig. 1. Polymers found in topologically non-trivial phase feature zero-energy edge states due to the bulk-boundary correspondence. Moreover, we also observe a significant renormalization of the electronic band gap for very short chains, while from a certain polymer length the energy band gap remains almost constant even in the phase transition, see Fig. 2D.

To get more insight into the underlying mechanism, we carried out total energy DFT calculations (41, 42) of free-standing pentacene-bridged polymers of various lengths. Fig. 4A displays evolution of the frontier's orbitals of polymers of different length forming the conduction and valence band in the limit of an infinite polymer. The red and blue color denotes bonding $\psi^b$ and antibonding $\psi^a$ character of given frontier orbital on the triple bond of the bridging unit, see Fig.

1. The phase transition occurs between polymers of 10 and 11 units and it is marked by the emergence of in-gap edge states (see black dots in Fig. 4A) and unification of the bonding (red) and antibonding (blue) character of empty and occupied states, respectively.

The spin restricted DFT calculations capture most of the experimental observations including (i) the presence of the phase transition accompanied by the avoided level crossing of frontier orbitals, see Fig. 4A; (ii) the transformation of the π-conjugation at a certain critical length (Fig. 4A), and (iii) the presence of the in-gap energy edge states as hallmarks of the topologically non-trivial phase, see fig. S5. In addition, calculated d$I$/d$V$ images (43) of single particle frontier HOMO/LUMO orbitals are in excellent agreement with the experimental evidence as shown on Fig. 2E. The only discrepancy we note is a shorter theoretical polymer length (~11 units) as compared to the experimental results in which the phase transition takes place. Such deviation could be attributed to limited precision of single particle DFT calculations as the precise position of the transition depends on employed exchange-correlation functional and/or absence of metallic substrate in our simulations. Despite this minor deficiency, we consider that the DFT calculations provide a reliable qualitative description of the underlying phase transition mechanism.

Interestingly, the frontier electronic orbitals involved in the avoided level crossing have an opposite bonding/antibonding character on the triple bond, see Figs. 1 and 4B. Therefore, the level crossing is directly related to the debilitation and elongation of the triple bond observed during the phase transition driven by the occupancy of anti/bonding frontier orbitals. This suggests that phonon softening could be responsible for the phase transition. Often spontaneous symmetry breaking in molecular systems with nondegenerate electronic states is caused by pseudo Jahn-Teller mechanism (38) driven by vibronic coupling of low-lying excited states to vibrational modes of particular symmetry to establish a new ground state at zero temperature. In next, we will show that indeed the pseudo Jahn-Teller is the driving mechanism of the phase transition in the pentacene polymers.

The total vibronic coupling $\gamma_Q^{tot}$ corresponding to a given soft phonon mode $Q$, which is a source of the phonon softening of the system, consist of sum of individual vibronic couplings $\gamma_Q^{i,j}$ between i,j frontier orbitals written as:

$$\gamma_Q^{ij} \sim \frac{\langle \psi_i | \partial H/\partial Q | \psi_j \rangle}{E_j - E_i} \quad (1)$$

, where $\psi_i, \psi_j, E_i, E_j$ are wavefunctions and eigen energies of i-th occupied and j-th unoccupied electronic states, respectively. Numerator $\langle \psi_i | \partial H/\partial Q | \psi_j \rangle$ represents off diagonal vibronic coupling constant (38) between the occupied and unoccupied electronic states mediated by vibrational mode $Q$. The presence of the energy difference $E_j - E_i$ in denominator in eq. 1 has two important consequences: (i) only low-energy excitation electronic states have significant contribution to the vibronic coupling; and (ii) a diminishing band gap enhances the vibronic coupling. The second consequence is often considered as a decisive factor in most of the spontaneous symmetry breaking processes. However, this argument cannot be fully adopted in our system as the experimental band gap remains almost constant near the critical point, as shown in Fig. 2D.

Thus, our findings indicate that the numerator in eq. 1 plays a decisive role in the phase transition. To understand the role of such term, we carried out an analysis of a pentacene dimer to identify a suitable coupling between vibrational modes and frontier orbitals involved in the pseudo Jahn-Teller (38, 44). Our calculations show that two highest occupied frontier orbitals (HOMO-1 and HOMO) have an opposite bonding $\psi^b$ and antibonding $\psi^a$ character on the triple bond of the bridging unit, respectively, and the same holds for the two lowest frontier unoccupied orbitals (LUMO and LUMO+1), as shown in Fig. 4B.

Moreover, we found that an in-plane stretching vibrational mode $Q$ of the triple bond in the bridging unit (see inset of Fig. 4C) has a suitable symmetry that allows significant vibronic coupling between the frontier electronic orbitals with the same bonding $\psi^b$ or antibonding $\psi^a$ symmetry. On the other hand, the term diminishes for two frontier orbitals of the opposite symmetry. Namely, in the case of the dimer chain shown in Fig. 4B we found that the vibronic coupling mediated by the stretching vibrational mode $Q$ between orbitals of the same symmetry is $\gamma_Q^{b,b} = \gamma_Q^{a,a}$ =0.73 eV/Å while for orbitals of different symmetries vanishes $\gamma_Q^{a,b} \sim 10^{-3}$. Tables S1-S3 summarize calculated vibronic coupling between occupied and unoccupied frontier orbitals for dimer, trimer and pentamer chain, whose total sum increases with the length of the chain. Thus, the longer polymer is, also the number of electronics states belonging to emerging valence and conduction bands (see Fig. 4A), as well as the number of the stretching vibrational modes Q of the bridge unit (see figs. S6-7) linearly increase. Consequently, the vibronic coupling steadily increases with the chain's longitude as shown on Fig. 4C till it reaches a critical value and the system undergoes a phase transition.

The phase transition from topologically trivial to non-trivial is accompanied by the crossing of the HOMO and LUMO frontier orbitals of distinct bonding character. This level crossing causes that the set of orbitals forming the valence band has the same antibonding character on the triple bond, while the conduction band is entirely formed by wavefunctions with the same bonding character, see Figs. 4A-B and figs. S5,8-9. Consequently, the vibronic coupling via the stretching mode vanishes. As the occupied orbitals of the valence band have antibonding symmetry, the character of the triple bond debilitates which causes the π-conjugation transformation from ethynylene-aromatic to cumulene-quinoid character accompanied by the topological transition. Our analysis shows that systems with conduction and valence bands composed by a mixture of wavefunctions of different symmetries are more prone to vibronic instabilities. We believe that this paradigm could be extended to other π-conjugated systems such as acenes or graphene nanoribbons (45).

Fig. S5 displays real-space wavefunctions of the pentacene polymer made of 15 units, which is found in topologically non-trivial phase featuring the in-gap edge states. Interestingly, we observe that the exchange bonding $\psi^b$ and antibonding $\psi^a$g orbitals take place mostly in the central part of the polymer, while at the edges the frontier orbitals preserve their original electronic character of short topologically trivial polymers. This finding matches well to experimental data acquired on topologically non-trivial polymers (see fig. S10) revealing a conservation of triple bond character on bridge units seen in high-resolution AFM images and reverse contrast of d$I$/d$V$ maps of HOMO/LUMO frontier orbitals at the ends of the polymers with respect to its central part (see Fig. 3E).

The extension of QPT to finite temperatures gives rise to the quantum criticality phenomenon, which is subject of an intensive research due to its exotic state of matter. There is a question if the quantum criticality may emerge also in the pentacene bridged polymer. Therefore, it would be interesting to see how the polymers of the critical lengths behave at elevated temperatures. Thus, we carried out QM/MM MD simulations (42, 46) of polymers made of 15 units (QM region) on the Au(111) substrate (MD region) at temperature of 100 Kelvin (for details see materials and methods). Interestingly, we observe cyclic reappearance of the in-gap edge states $\psi^e$ (black) accompanied by level crossing of bonding $\psi^a$ (red) and antibonding $\psi^b$ (blue) frontier orbitals in energy, as shown in Fig. 4D and supplementary video S1. This indicates continuous fluctuation of the polymer between two topological phases at finite temperature. Such fluctuations are driven by presence of a new stable vibrational mode with time oscillation period of ~1ps. This vibrational mode has a character of a breathing mode continuously varying the total length of the polymer. On the contrary, the level crossing and cyclic reappearance of the edge state is completely absent in MD simulations of a topologically trivial chain made of 5 units, see fig. S11. Despite we are aware of the limitations of the correct description of the electronic structure using the DFT method, we consider that the coherent evolvement on time scale of at least tens of picoseconds between the two topological phases may indicate the existence of the quantum criticality in π-conjugated polymers located near the topological phase transition.

We believe that these findings will motivate further research with appropriated experimental techniques. Unfortunately, such quantum coherent evolvement cannot be directly inspected by low-temperature microscopy, which operation at elevated temperatures causes significant loss of the spatial and energy resolution. Instead, for example conductance measurements through single chains (47, 48) near the quantum phase transition could explore the possible critical behavior the conductivity of the polymers at finite temperature.

We anticipate that our results could give rise to fluctuating quasi-metallic entangled phase within the polymer. We envision our observations will stimulate more elaborated studies of the electronic structure of π-conjugated polymers and the role of the size dimensions of such polymers regarding topological quantum classes and emergence of complex quantum ground states in the quantum criticality regime.

In addition, our new recipe based on extra hydrogenation and atomic manipulation provides avenues to engineer polymer segments with different topological phases, or two separated segments of the same topology, whereby to study quantum phenomena emerging from trivial/non-trivial interfaces, as well as the interaction of two topologically protected edge states and their protection from disorder or external stimuli.

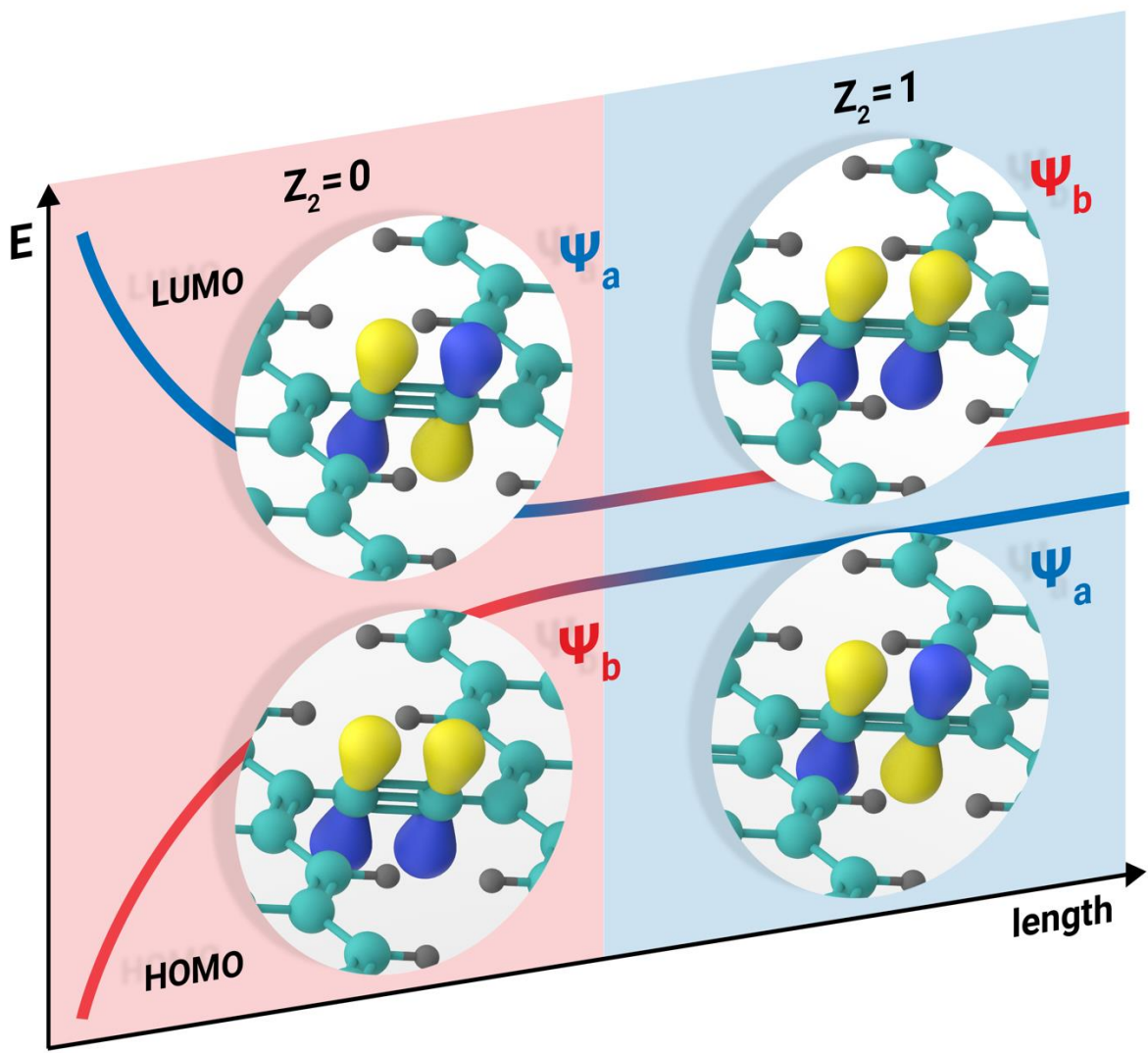

**Fig. 1 Schematic view of the topological quantum phase transition in π-conjugated pentacene bridged polymers with their length.** The phase transition is driven by avoid level crossing of two frontier HOMO/LUMO orbitals with distinct bonding ($\psi^b$) and antibonding ($\psi^a$) character of π-orbitals located on two carbon atoms forming the bridge. This exchange of two frontier bonding $\psi^b$ and antibonding $\psi^a$ orbitals in the avoid level crossing debilitates the triple bond in the bridge causing transformation of the π-resonant form of the polymer and emergence of the topologically nontrivial quantum phase in-gap edge states.

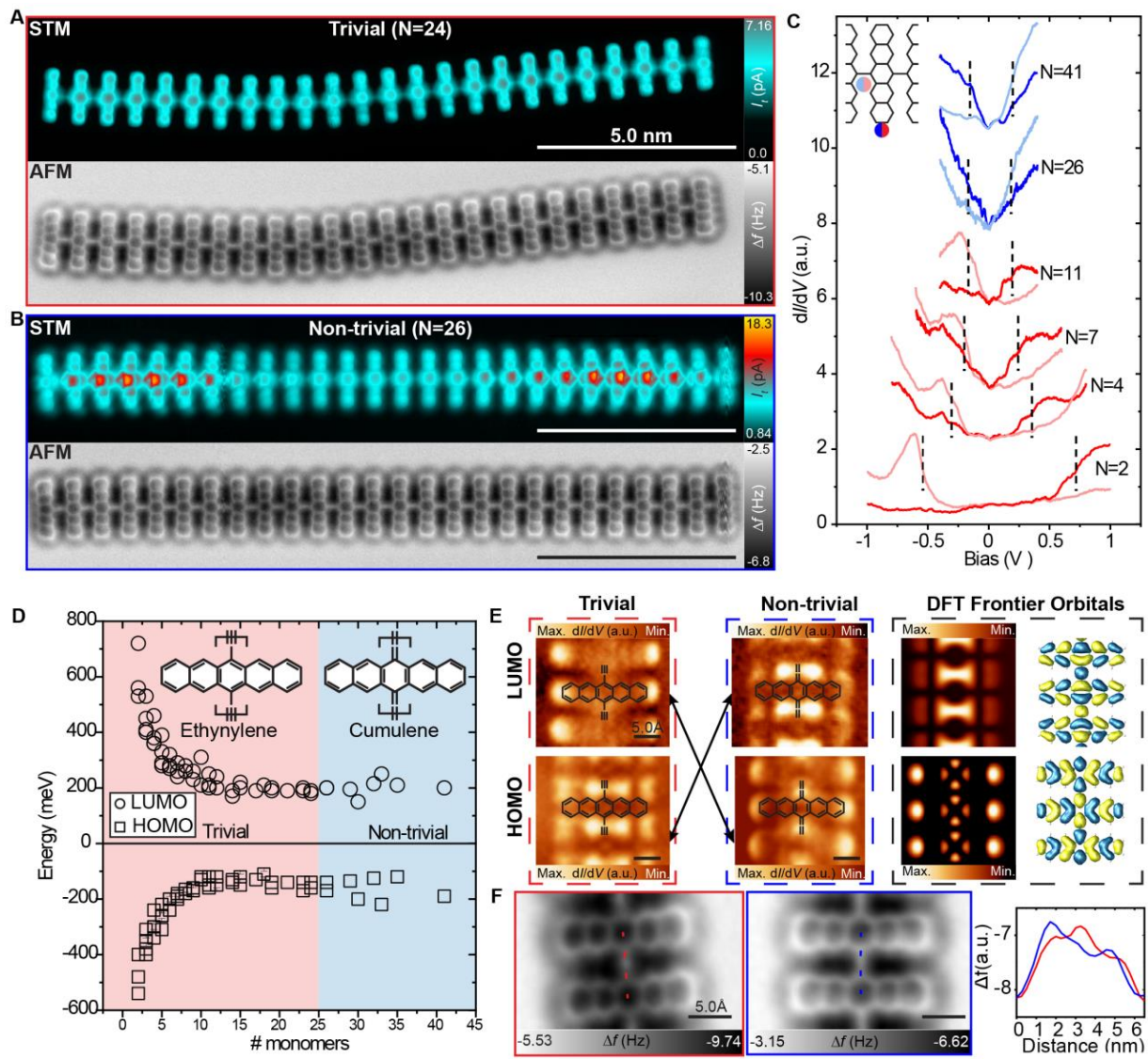

**Fig. 2 Length dependent characterization of pentacene polymers.** (**A**) Constant height STM (top) and nc-AFM (bottom) image of a 24 monomers pentacene. (**B**) Constant height STM (top) and nc-AFM (bottom) image of a 26 monomers pentacene. (**C**) d$I$/d$V$ spectra of polymers of different lengths acquired. Red/blue color of curves corresponds to topologically trivial/non-trivial chains. Inset: schematic view of the positions where the d$I$/d$V$ curves were acquired within a pentacene unit in central part of the polymer. (**D**) HOMO/LUMO energy evolution with the polymer length. (**E**) Experimental d$I$/d$V$ maps of the LUMO (top) and HOMO (down) of trivial and non-trivial chains (left and center respectively). Simulated d$I$/d$V$ maps of the frontier orbitals (right) of the infinite pentacene bridged chain. (**F**) Zoom in the marked red/blue rectangles in AFM images from panels (A-B). Profile along the red/blue line in the zoomed images. The profile displays a bright spot as a peak in red line associated with the triple bond in the bridge unit, which is missing in the blue line indicating a double bond between two carbon atoms of the bridge unit.

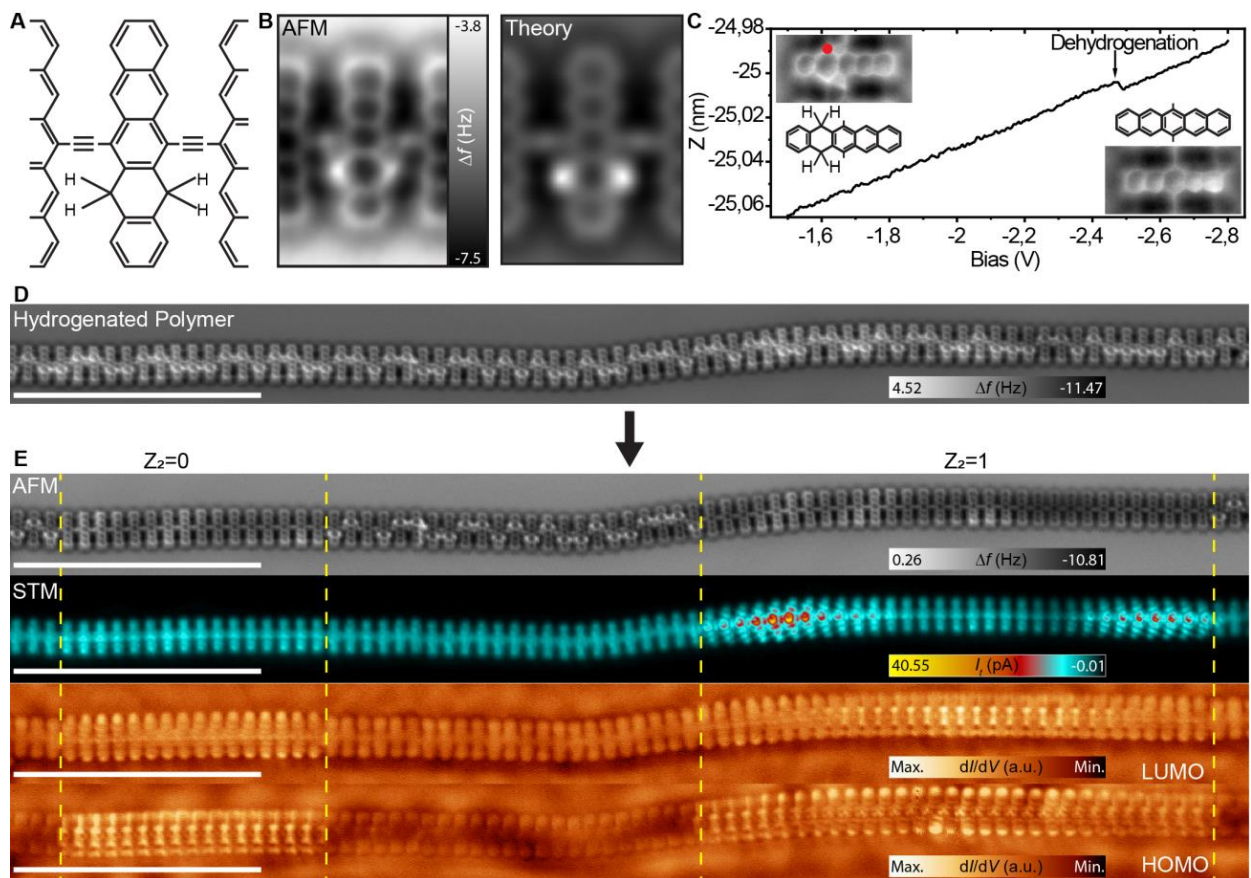

**Fig. 3 Engineering pentacene polymers by tip induced dehydrogenation.** (**A**) Chemical sketch of a hydrogenated pentacene polymer. (**B**) Experimental (left) and simulated (right) AFM images of the hydrogenated pentacene polymer. (**C**) Characteristic distance vs. bias curve acquired at the position marked by the red dot to remove two extra hydrogens form the pentacene unit. Inset top left, a nc-AFM image of the pentacene unit before the curve was taken and corresponding chemical sketch. Inset down right, a nc-AFM image after the removal of two extra hydrogen atoms and corresponding chemical sketch. (**D**) nc-AFM image of a fully hydrogenated pentacene polymer. (**E**) From top to bottom: nc-AFM image of the chain showed in panel (D) after thel hydrogen manipulations at two distinct part of the polymer to produce a short topologically trivial (left) and a long topologically non-trivial (right) segments of the polymer; low bias STM image of the polymer chain; LUMO and HOMO d$I$/d$V$ maps of the engineered polymer revealing the exchange of frontier orbitals between topologically trivial and non-trivial segments of the polymer. Dashed yellow lines added to guide the eye.

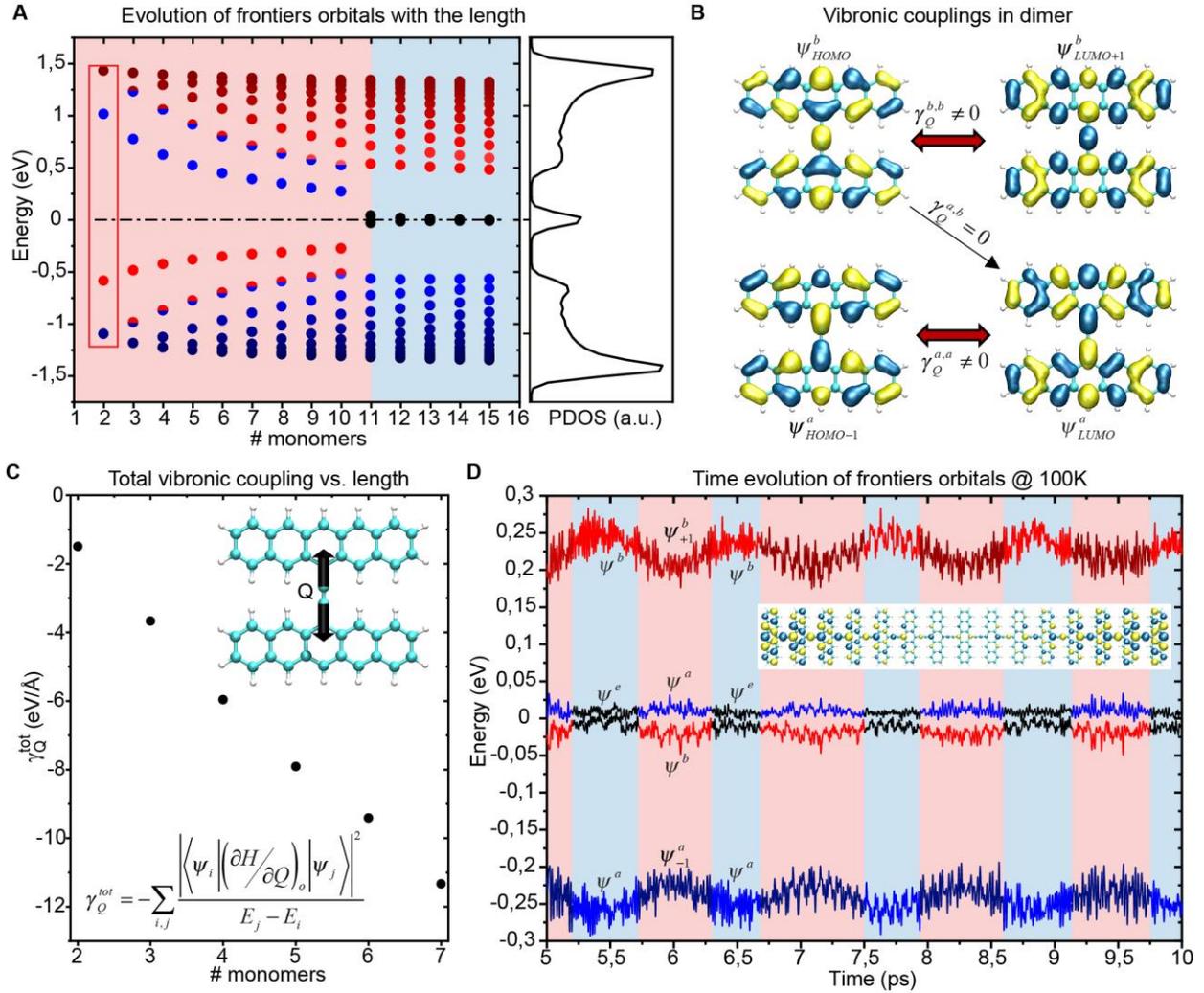

**Fig. 4 Theoretical simulations of electronic states and vibronic couplings of pentacene bridged polymers.** (**A**) Calculated electronic spectra of frontier orbitals of the pentacene polymer made of different monomers and corresponding projected density of state (right) for infinite polymer. The red and blue circles denote bonding $\psi^b$ and antibonding $\psi^a$ character of given frontier orbital on the triple bond of the bridging unit., similarly black circles represent the in-gap edge states emerging in the topologically non-trivial quantum phase. (**B**) Two highest occupied (left) and two lowest unoccupied (right) frontier orbitals of the dimer (see box in (A)) featuring different bonding $\psi^b$ and antibonding $\psi^a$ character and corresponding strength of vibronic coupling between orbitals of different bonding character, $\gamma_Q^{b,b} = \gamma_Q^{a,a} \neq 0$ and $\gamma_Q^{a,b} \sim 0$ (see table S1). Strength of vibronic coupling between frontier orbitals is also schematically represented by thickness of corresponding arrows between orbitals. (**C**) Evolution of total vibronic coupling $\gamma_Q^{tot}$ with the length of the polymer. Inset shows the in-plane stretching vibrational mode which mediates the vibronic coupling between frontier orbitals. (**D**) Time evolution of selected electronic states of the polymer made of 15 units at 100 Kelvin showing reappearance of the in-gap edge states $\psi_e$ (black) accompanied by level crossing of bonding $\psi^a$ (red) and antibonding $\psi^b$ (blue) frontier orbitals in energy, inset figure shows the electronic wavefunction of the in-gap edge state.

**Materials and Methods**

Experimental Methods

All the experiments were carried out in an ultra-high vacuum chamber hosting a low temperature (4.2 K) STM/AFM (Createc). Bias voltage is applied to the sample. Base pressure was ~5.0 E-10 mbar.

STM/AFM images were acquired with a Pt/Ir tip mounted on a qPlus sensor (resonant frequency of ~30 kHz; stiffness of ~1,800 N m–1), cut and sharpened by focused ion beam. The high resolution nc-AFM images were obtained at constant-heigh, operating the qPlus sensor in the frequency modulation mode using a constant amplitude oscillation of 50 pm. The metal tip apex was prepared by gentle indentation (~ 1 nm) on the clean Au(111) substrate and functionalized with a CO molecule dosed at low-temperature.

Conductance dI/dV spectra and maps were taken using a lock-in technique, with an ac voltage (frequency: 700-800 Hz, amplitude: 1-10 mV rms) added to the dc sample bias. dI/dV maps were recorded in constant current mode.

The data were processed using the WSxM software (49). All images and spectroscopy plots presented corresponds to raw data with only slope and gaussian corrections.

Au(111) substrate was prepared by standard $Ar^+$ sputtering at energies from 1.0 to 1.5 kV and annealing at 800 K afterwards.

The 4BrPn molecules were deposited from a tantalum crucible at ~450 K on the fresh cleaned Au(111) crystal held at RT.

Hydrogenation of the pentacene polymers was achieved by placing the sample out from the cryostat and increasing the molecular hydrogen partial pressure of the STM chamber while keeping the ion gauge filament on. The molecular hydrogen in the residual gas was then cracked by the filament and adsorbed by our polymers. Control experiments where the polymer-decorated sample was left out with the ion gauge off showed no hydrogenation of the chains even after long periods of time.

Theoretical Methods

Our QM/MM simulations were performed using Fireball/Amber method, (46) that combines interface force field (50) for the Au surface with Fireball local orbital DFT (42) for the molecule. Fireball calculations used the BLYP exchange-correlation functional (51, 52) with D3 corrections (53) and norm-conserving pseudopotentials, with a basis set of optimized numerical atomic-like orbitals with cut off radii H (s=5.42 a.u.) and C (s,p=5.95 a.u.) (54). The QM/MM MD simulation for the 5, 15 and 20 chains were performed at 100K with a langevin thermostat for 10 ps with a time step of 0.5 fs. Vibronic couplings were calculated using formalism for calculations of non-adiabatic couplings in numerical local basis set described in (44) using Fireball code.

For all free-standing finite and infinite chains, DFT calculations were done using Fireball code with parameters described above. We also obtained very similar results using FHI-AIMS (41), where the geometry optimizations and electronic structure analyses were performed using B3LYP (55) exchange–correlation functional. Systems were allowed to relax until the remaining atomic forces reached below 10-2 eV Å-1. To sample the Brillouin zone, for the infinite system with a periodic boundary condition, a Monkhorst–Pack grid of $18 \times 1 \times 1$ was used and for the finite systems, one k-point at the Gamma point. Theoretical dI/dV maps were calculated from electronic structure obtained from the DFT Fireball (42) program package and with Probe Particle Scanning Probe Microscopy (PP-SPM) code for an s-like orbital tip (43, 56). The AFM simulations of the hydrogenated pentacene were performed by PP-SPM code using a DFT-relaxed atomic structure

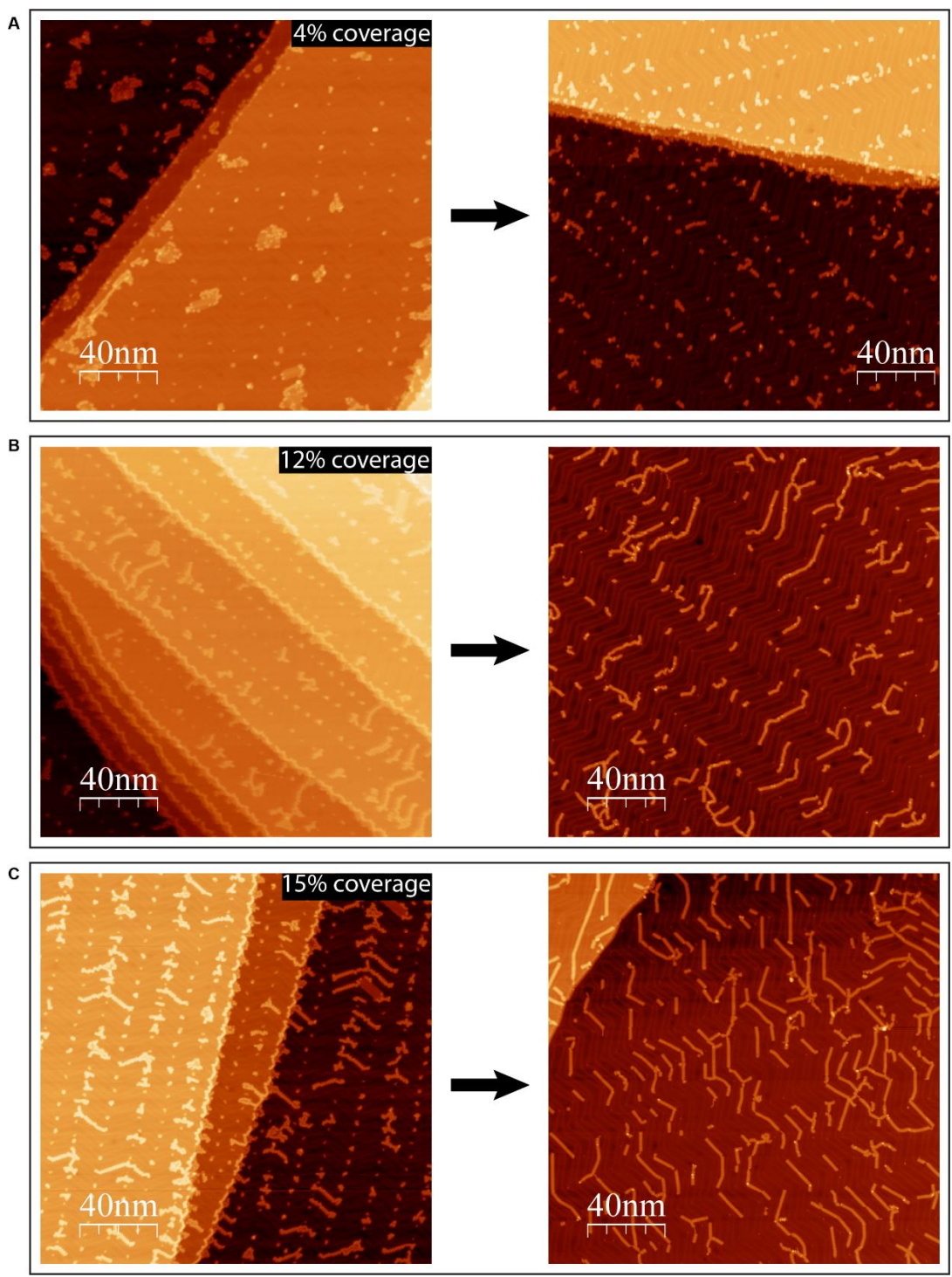

**Fig. S1. Coverage/polymer chain length dependance.** (**A**) STM image of a 4% coverage 4BrPn on Au(111) (left) and the polymer chains obtained after annealing the sample at 500K (right). (**B**) STM image of a 12% coverage 4BrPn on Au(111) (left) and the polymer chains obtained after annealing the sample at 500K (right). (**C**) STM image of a 15% coverage 4BrPn on Au(111) (left) and the polymer chains obtained after annealing the sample at 500K (right).

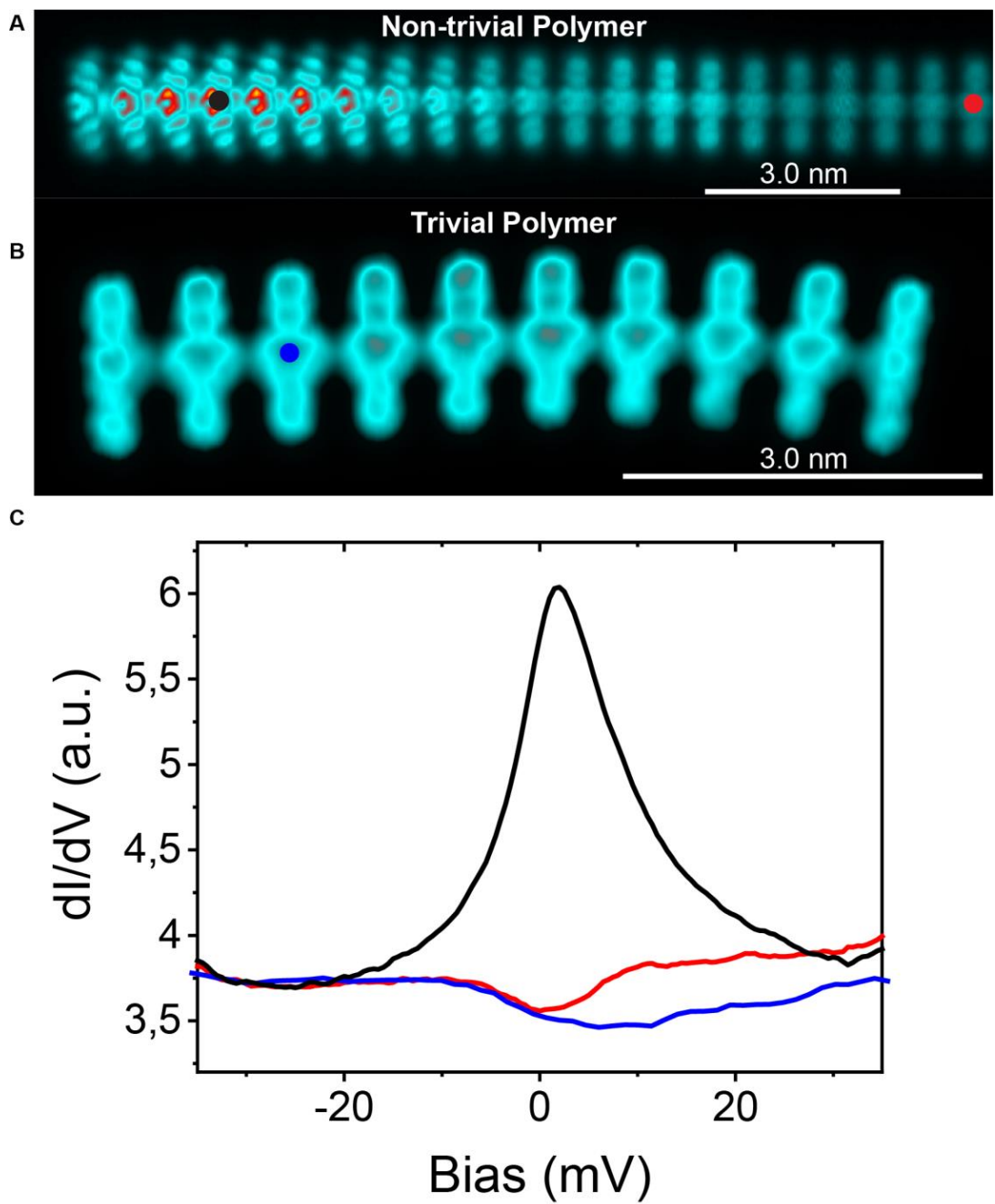

**Fig. S2. Trivial / non-trivial spectroscopy.** (**A**) Constant height STM image of a non-trivial chain's edge. (**B**) Constant height STM image of a 10 monomers trivial chain. (**C**) dI/dV curves obtained in the dots marked in the constant height STM images above.

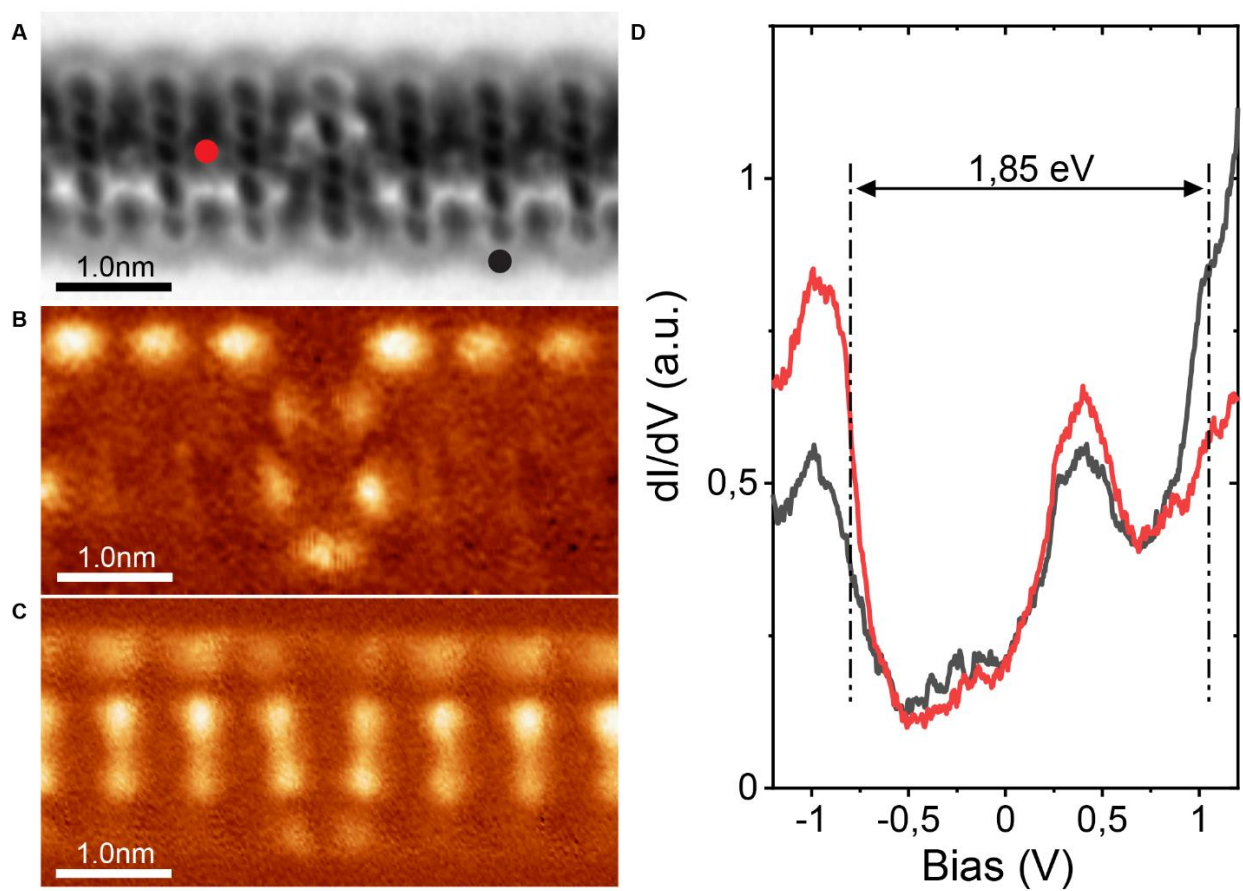

**Fig. S3. Hydrogenated polymers gap.** (**A**) Constant heigh AFM image of a hydrogenated chain (**B**) d*I*/d*V* map at -830 mV of the same polymer chains as (A). (**C**) dI/dV map at -950 mV of the same polymer chain as (A). (**D**) dI/dV spectra acquired in the red and black dots marked in panel (A).

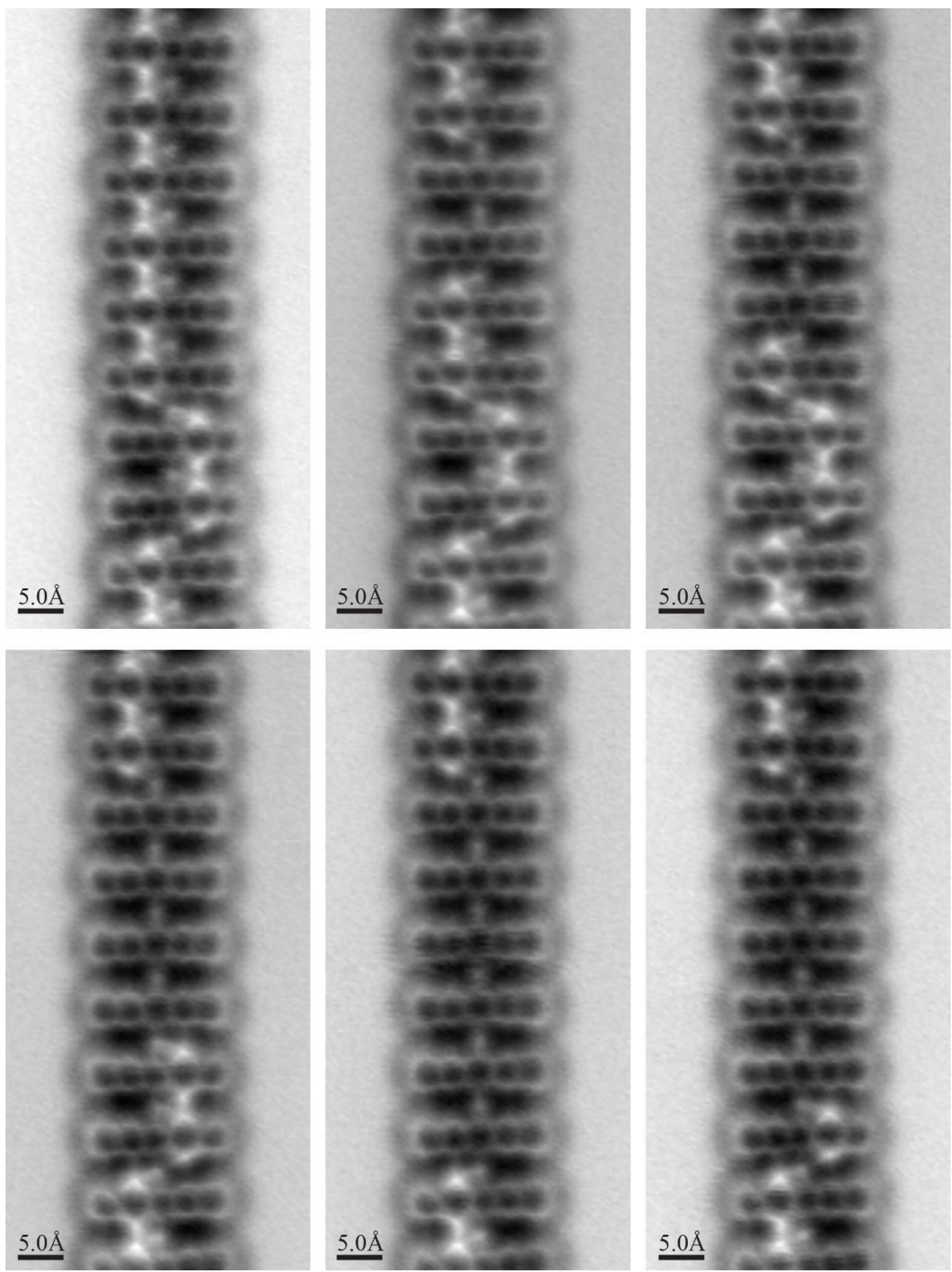

**Fig. S4. Hydrogen manipulation in hydrogenated polymers.** From left to right and top to bottom, constant height nc-AFM images of the same region of a polymer after consecutive hydrogen removal has been done.

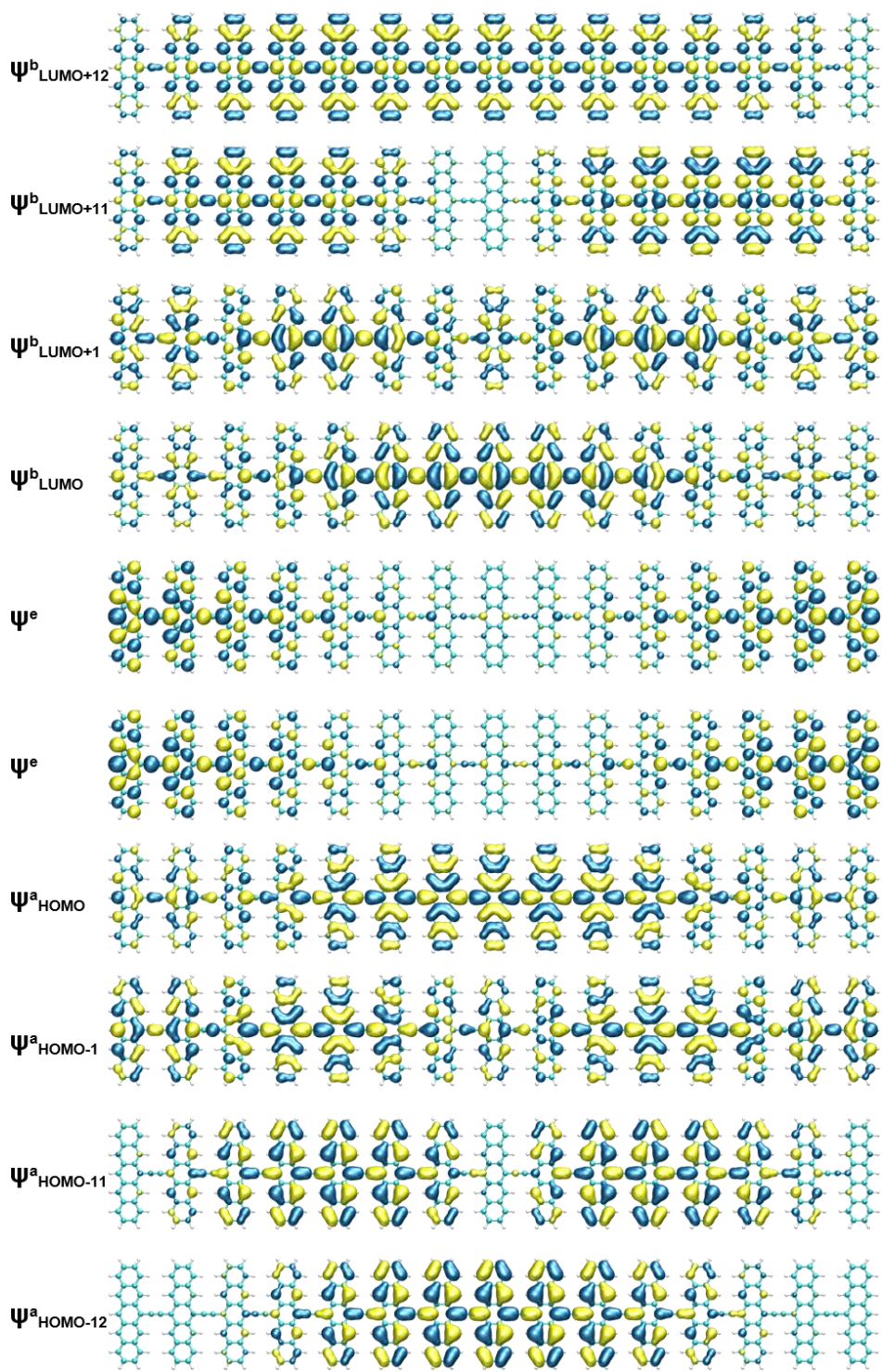

**Fig. S5. Frontier orbitals for topologically non-trivial pentacene chain made of 15 units.** Selected occupied (HOMO) and unoccupied (LUMO) frontiers orbitals $\psi$ belonging to emerging valence and conduction bands (see Figure 4a) obtained from gas phase DFT calculation. Real space images reveal bonding $\psi^b$ and antibonding $\psi^b$ character of $\pi$-bonds on the bridging units of individual frontiers orbitals and presence of two in-gap edge states $\psi^e$.

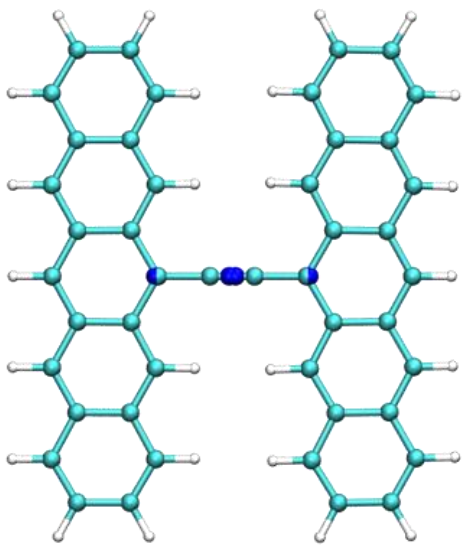

**Fig. S6. Calculated vibrational stretching mode causing the pseudo JT effect for the dimer chain.** Normal mode $Q$ associated with the stretching of the bridge for a dimer chain. In dark blue the atoms in the maximum amplitude.

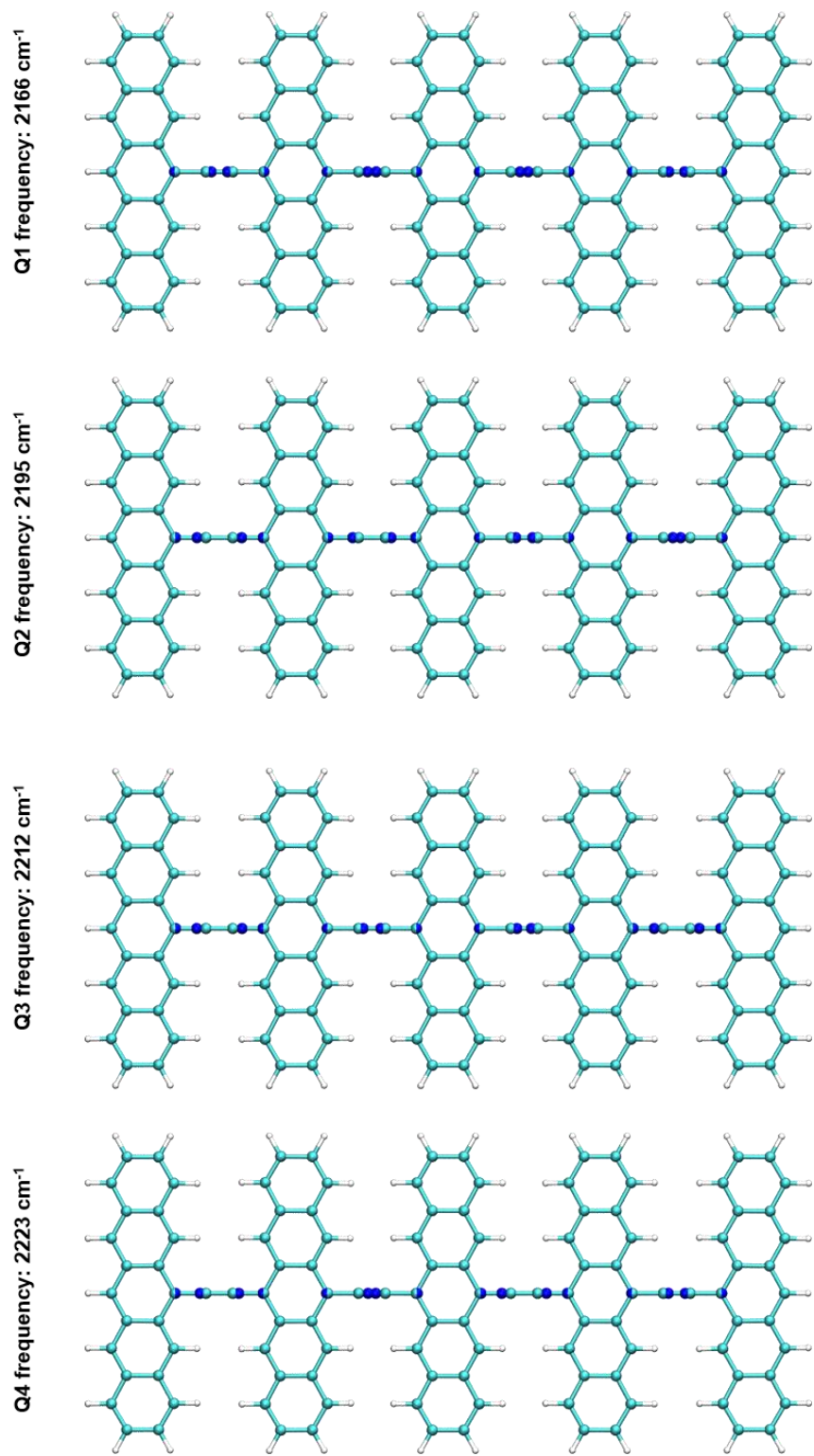

**Fig. S7. Calculated vibrational stretching mode causing the pseudo JT effect for the pentamer chain.** Four normal modes $Q$ associated with the stretching of the bridge for the pentamer chain. In dark blue the atoms in the maximum amplitude.

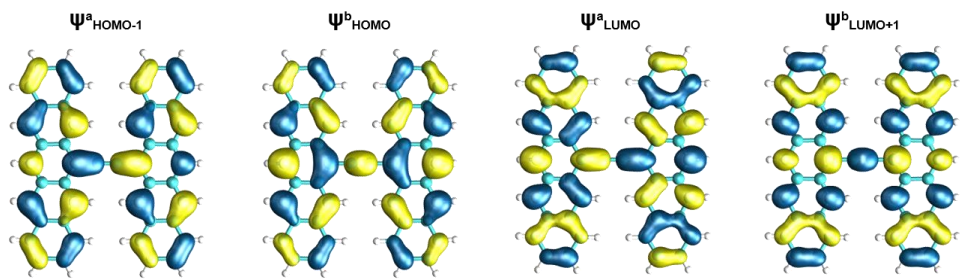

**Fig. S8. Frontier orbitals for topologically trivial dimer pentacene chain.** Calculated occupied (HOMO) and unoccupied (LUMO) frontiers orbitals $\psi$ belonging to emerging valence and conduction bands (see Figure 4a) obtained from gas phase DFT calculation. Real space images reveal bonding $\psi^b$ and antibonding $\psi^b$ character of $\pi$-bonds on the bridging unit of individual frontiers orbitals.

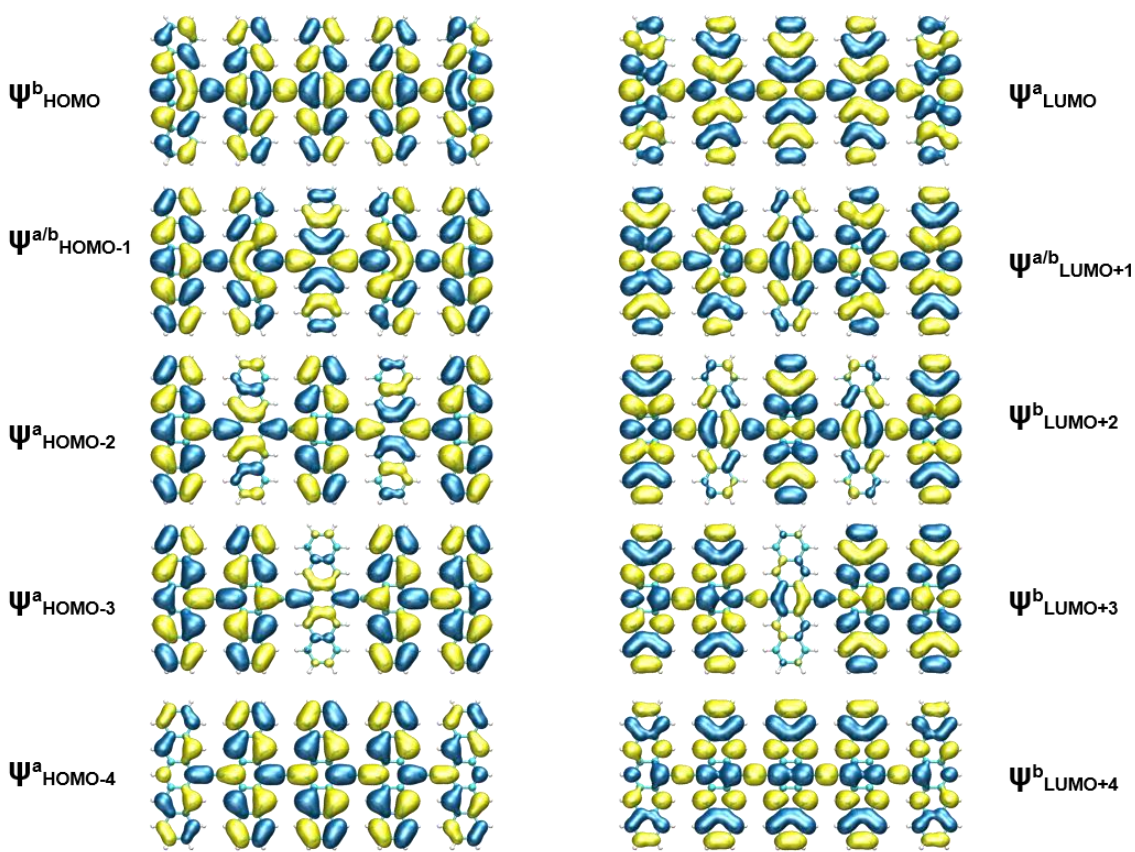

**Fig. S9. Frontier orbitals for topologically trivial pentamer pentacene chain.** Calculated occupied (HOMO) and unoccupied (LUMO) frontiers orbitals $\psi$ belonging to emerging valence and conduction bands (see Figure 4a) obtained from gas phase DFT calculation. Real space images reveal bonding $\psi^b$ and antibonding $\psi^b$ character of $\pi$-bonds on the bridging unit of individual frontiers orbitals.

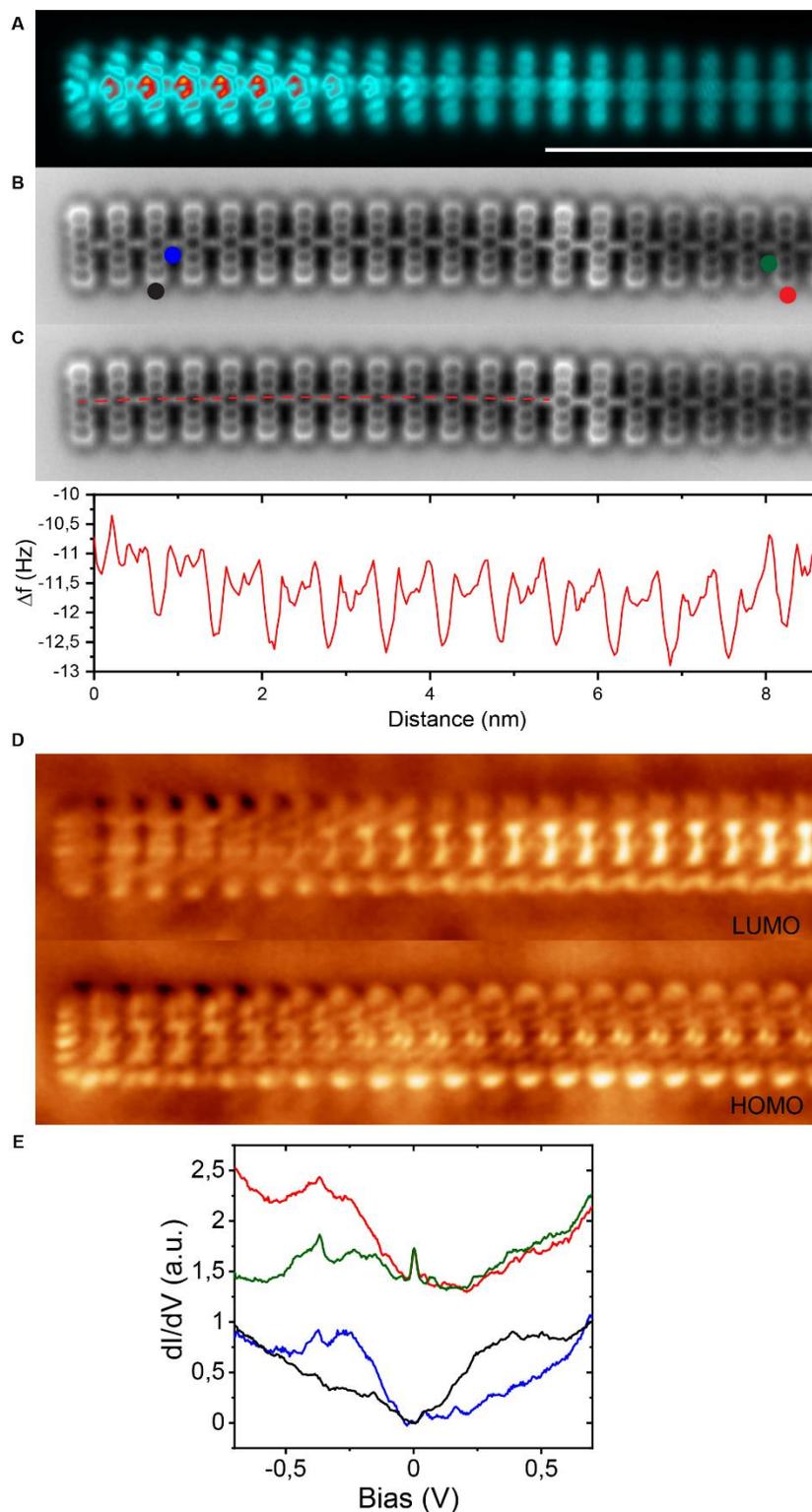

**Fig. S10. Band inversion along non-trivial chains.** (**A**) Constant heigh STM image of the 20 first monomers of a non-trivial chain. (**B**) Simultaneous constant heigh AFM. (**C**) AFM image with profile along the line marked in red. (**D**)d$I$/d$V$ map at 200mV (LUMO) and -190mV (HOMO) acquired in the same polymer. (**E**) d$I$/d$V$ spectra acquired in the dots marked in panel b) showing the inversion of the bands.

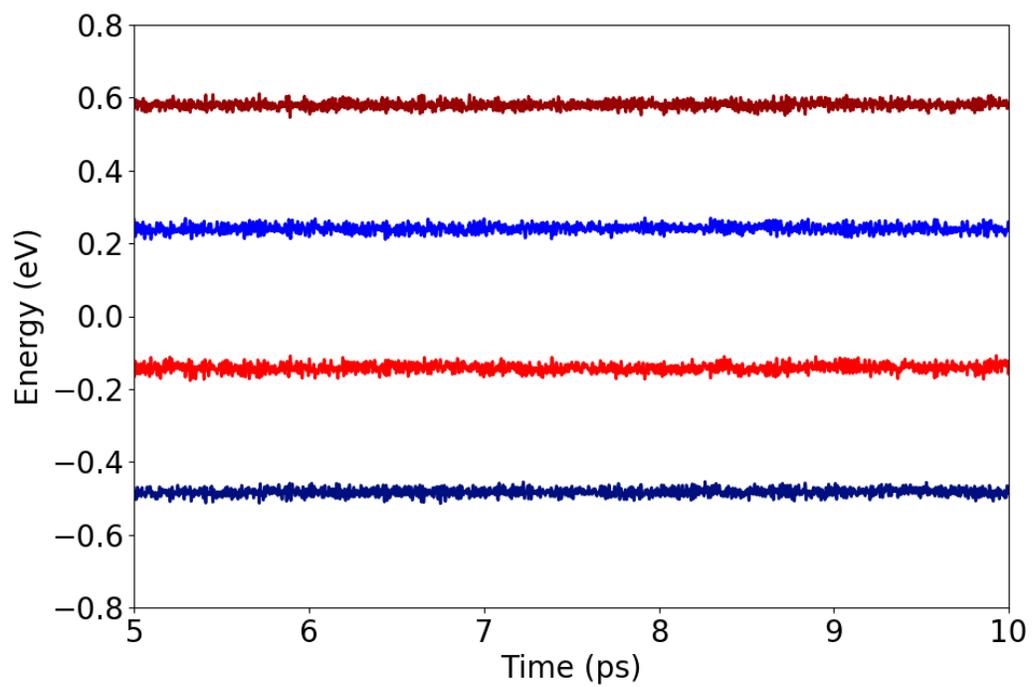

**Fig. S11.** Time evolution of HOMO-1 (dark blue), HOMO (red), LUMO (blue) and LUMO+1 (dark red) for a 5-unit pentacene.

| $Q_1$ | $\psi_{LUMO}$ | $\psi_{LUMO+1}$ |
|---|---|---|
| $\psi_{HOMO}$ | -0.015 | -0.731 |
| $\psi_{HOMO-1}$ | -0.731 | -0.008 |
| $\sum \gamma_{ij}^{Q_2}$ | | 1.485 |

**Table S1. Overview of the calculated vibronic couplings for the dimer pentacene oligomer.** Calculated values of individual vibronic couplings $\gamma_{ij}^Q$ [eV/Å] between occupied and unoccupied frontiers orbitals $\psi$ belonging to set of emerging valence and conduction bands for the dimer pentacene polymer (see fig. S8) of the stretching mode $Q$ (see fig. S6).

| $Q_1$ | $\psi_{LUMO}$ | $\psi_{LUMO+1}$ | $\psi_{LUMO+2}$ |
|---|---|---|---|
| $\psi_{HOMO}$ | -0.019 | -0.465 | -0.371 |
| $\psi_{HOMO-1}$ | -0.483 | -0.003 | -0.084 |
| $\psi_{HOMO-2}$ | -0.341 | -0.071 | -0.004 |
| | | $\sum \gamma_{ij}^{Q_1}$ | **-1.841** |

| $Q_2$ | $\psi_{LUMO}$ | $\psi_{LUMO+1}$ | $\psi_{LUMO+2}$ |
|---|---|---|---|
| $\psi_{HOMO}$ | -0.016 | -0.46 | -0.366 |
| $\psi_{HOMO-1}$ | -0.485 | -0.003 | -0.086 |
| $\psi_{HOMO-2}$ | -0.33 | -0.071 | -0.004 |
| | | $\sum \gamma_{ij}^{Q_2}$ | **-1.821** |

**Table S2. Overview of the calculated vibronic couplings for the trimer pentacene oligomer.** Calculated values of individual vibronic couplings $\gamma_{ij}^Q$ [eV/Å] between occupied and unoccupied frontiers orbitals $\psi$ belonging to set of emerging valence and conduction bands for the trimer pentacene polymer of two stretching modes $Q_{1-2}$.

| $Q_1$ | $\psi_{LUMO}$ | $\psi_{LUMO+1}$ | $\psi_{LUMO+2}$ | $\psi_{LUMO+3}$ | $\psi_{LUMO+4}$ |
|---|---|---|---|---|---|
| $\psi_{HOMO}$ | -0.01 | -0.409 | -0.153 | -0.088 | -0.195 |
| $\psi_{HOMO-1}$ | -0.401 | -0.003 | -0.101 | -0.013 | -0.035 |
| $\psi_{HOMO-2}$ | -0.147 | -0.107 | -0.001 | -0.031 | -0.022 |
| $\psi_{HOMO-3}$ | -0.092 | -0.012 | -0.031 | 0 | -0.013 |
| $\psi_{HOMO-4}$ | -0.191 | -0.028 | -0.021 | -0.011 | -0.002 |
| | | | | $\sum \gamma_{ij}^{Q_1}$ | **-2.117** |

| $Q_2$ | $\psi_{LUMO}$ | $\psi_{LUMO+1}$ | $\psi_{LUMO+2}$ | $\psi_{LUMO+3}$ | $\psi_{LUMO+4}$ |
|---|---|---|---|---|---|
| $\psi_{HOMO}$ | -0.011 | -0.384 | -0.159 | -0.098 | -0.185 |
| $\psi_{HOMO-1}$ | -0.383 | -0.002 | -0.1 | -0.024 | -0.038 |
| $\psi_{HOMO-2}$ | -0.156 | -0.106 | -0.001 | -0.031 | -0.022 |
| $\psi_{HOMO-3}$ | -0.101 | -0.023 | -0.031 | -0.001 | -0.013 |
| $\psi_{HOMO-4}$ | -0.183 | -0.031 | -0.02 | -0.01 | -0.002 |
| | | | | $\sum \gamma_{ij}^{Q_2}$ | **-2.115** |

| $Q_3$ | $\psi_{LUMO}$ | $\psi_{LUMO+1}$ | $\psi_{LUMO+2}$ | $\psi_{LUMO+3}$ | $\psi_{LUMO+4}$ |
|---|---|---|---|---|---|
| $\psi_{HOMO}$ | -0.014 | -0.12 | -0.209 | -0.186 | -0.084 |
| $\psi_{HOMO-1}$ | -0.177 | -0.002 | -0.082 | -0.126 | -0.059 |
| $\psi_{HOMO-2}$ | -0.228 | -0.084 | -0.001 | -0.031 | -0.022 |
| $\psi_{HOMO-3}$ | -0.183 | -0.12 | -0.025 | -0.003 | -0.005 |
| $\psi_{HOMO-4}$ | -0.085 | -0.058 | -0.018 | -0.003 | -0.001 |
| | | | | $\sum \gamma_{ij}^{Q_3}$ | **-1.926** |

| $Q_4$ | $\psi_{LUMO}$ | $\psi_{LUMO+1}$ | $\psi_{LUMO+2}$ | $\psi_{LUMO+3}$ | $\psi_{LUMO+4}$ |
|---|---|---|---|---|---|
| $\psi_{HOMO}$ | -0.013 | -0.123 | -0.207 | -0.181 | -0.088 |
| $\psi_{HOMO-1}$ | -0.18 | -0.002 | -0.082 | -0.123 | -0.061 |
| $\psi_{HOMO-2}$ | -0.226 | -0.084 | -0.001 | -0.03 | -0.023 |
| $\psi_{HOMO-3}$ | -0.179 | -0.116 | -0.024 | -0.003 | -0.005 |
| $\psi_{HOMO-4}$ | -0.089 | -0.06 | -0.019 | -0.003 | -0.001 |
| | | | | $\sum \gamma_{ij}^{Q_4}$ | **-1.923** |

**Table S3. Overview of the calculated vibronic couplings for the pentamer pentacene oligomer.** Calculated values of individual vibronic couplings $\gamma_{ij}^Q$ [eV/Å] between occupied and unoccupied frontiers orbitals $\psi$ belonging to set of emerging valence and conduction bands for the pentamer pentacene polymer (see fig. S9) of four stretching modes $Q_{1-4}$ (see Fig. S7).